\documentclass[apl,prb,showpacs,twocolumn,amsmath,amssymb,superscriptaddress,affilletter]{revtex4-1}
\usepackage{color}
\usepackage{graphicx}
\usepackage{tabularx}
\usepackage[T1]{fontenc}
\usepackage[cp1250]{inputenc}
\usepackage{dcolumn}
\usepackage{amsmath,amsfonts,amssymb}

\newcommand{\Xm}{X$^-$}
\newcommand{\Xp}{X$^+$}

\begin{document}

\title{Tuning the inter-shell splitting in self-assembled CdTe quantum dots}


\author{K.~Kukliński}   \affiliation{Institute of Physics, Polish Academy of Sciences, Al. Lotników 32/46, 02-668 Warsaw, Poland}

\author{Ł.~Kłopotowski}
\email[Corresponding author: ]{lukasz.klopotowski@ifpan.edu.pl}
\affiliation{Institute of Physics, Polish Academy of Sciences, Al. Lotników 32/46, 02-668 Warsaw, Poland}

\author{K.~Fronc}       \affiliation{Institute of Physics, Polish Academy of Sciences, Al. Lotników 32/46, 02-668 Warsaw, Poland}

\author{M.~Wiater}      \affiliation{Institute of Physics, Polish Academy of Sciences, Al. Lotników 32/46, 02-668 Warsaw, Poland}

\author{P.~Wojnar}      \affiliation{Institute of Physics, Polish Academy of Sciences, Al. Lotników 32/46, 02-668 Warsaw, Poland}

\author{P.~Rutkowski}   \affiliation{Institute of Physics, Polish Academy of Sciences, Al. Lotników 32/46, 02-668 Warsaw, Poland}

\author{V.~Voliotis}    \affiliation{Institut des NanoSciences de Paris, Université Pierre et Marie Curie, CNRS, 4 place Jussieu, 75252 Paris, France}

\author{R.~Grousson}     \affiliation{Institut des NanoSciences de Paris, Université Pierre et Marie Curie, CNRS, 4 place Jussieu, 75252 Paris, France}

\author{G.~Karczewski}  \affiliation{Institute of Physics, Polish Academy of Sciences, Al. Lotników 32/46, 02-668 Warsaw, Poland}

\author{J.~Kossut}      \affiliation{Institute of Physics, Polish Academy of Sciences, Al. Lotników 32/46, 02-668 Warsaw, Poland}

\author{T.~Wojtowicz}   \affiliation{Institute of Physics, Polish Academy of Sciences, Al. Lotników 32/46, 02-668 Warsaw, Poland}

\date{\today}

\begin{abstract}
We present photoluminescence studies of highly excited single self--assembled CdTe quantum dots under continuous--wave and pulsed excitations. We observe appearance of emission bands related to sequential filling of s--, p-- and d--shells. We analyze the inter-shell splitting for five samples, in which the dots were formed from a strained CdTe layer of different width. We find that with increasing the CdTe layer width the inter-shell splitting increases. In a time resolved measurement, we observe a radiative cascade between transitions involving one, two, and more than two excitons.

\end{abstract}

\pacs{78.67.Hc, 72.20.Jv, 71.55.Gs}%

\maketitle

Quantum dots (QDs) are often referred to as artificial atoms due to their zero-dimensional density of states and a shell-like structure of energy levels \cite{jac98}. Epitaxial, self-assembled QDs are usually strongly flattened along the growth axis, yielding a separation of the in-plane and vertical degrees of freedom. The quantization of the in-plane motion results in shells with lowered degeneracies with respect to spherically symmetric atoms and nanocrystals, albeit wave functions associated with subsequent shells retain the symmetries of the orbital momentum eigenfunctions. In order to implement QDs into one of the many envisioned future spin-based devices,\cite{mic09} a control over their morphology, and thus energy structure, is essential. In particular, engineering of the inter-level spacing is important, since mixing of different orbital states strongly accelerates spin relaxation. \cite{kha00} On the other hand, tailoring of the ground state photoluminescence (PL) energy is crucial for non-classical light sources \cite{mic00} to match the QD emission with the cavity modes of photonic structures.
Past work on size and shape engineering and thus tuning of the optical properties involved rapid thermal annealing (RTA),\cite{faf99,all01,mac03,woj07}  controlling the growth temperature\cite{faf99prb}, indium flush technique\cite{faf99prb,was99,faf99lasing} or selective etching and overgrowth of nanoholes. \cite{wan09} In particular, it was shown that in the InGaAs system, the inter-shell spacings can be increased by applying a lower growth temperature\cite{faf99}, by executing the indium flush after deposition of a thicker GaAs layer\cite{faf99lasing} or by producing a deeper nanohole -- conversely a lower QD.\cite{wan09} In this work, we study epitaxial CdTe QDs with ZnTe barriers for which no comprehensive study of inter-shell spacings is available. These dots are grown using a modified Stranski-Krastanov procedure.\cite{tin03} Namely, onto a strained 2D layer of CdTe, a layer of amorphous tellurium is deposited to change the balance between the elastic and surface energies, and the dots are formed upon desorption of the tellurium. We find that the splitting between the s-- and p--shells is proportional to the width of the CdTe layer providing a new tool for controlling the in-plane confinement in QDs. Moreover, in a time-resolved PL measurement we demonstrate a radiative cascade between a p--shell and s--shell and reproduce the PL transients with a rate equation model.

The samples were grown by molecular beam epitaxy on a (100)-oriented GaAs substrate. First, a 4 $\mu$m thick CdTe buffer layer was deposited. Then, a 1.4 $\mu$m ZnTe barrier layer was grown. Next, a layer of strained CdTe was deposited with thickness varying from 2 to 6 monolayers (ML) i.e.\ from 6.5 to 19.5 \AA. The dots were formed from this layer using the tellurium desorption procedure\cite{tin03} mentioned above. Finally, the dot layer was covered with 50 nm thick ZnTe barrier layer.

We measured PL of single dots excited with a continuous wave (cw) or a pulsed laser. The excitation wavelength in both cases was 532 nm i.\ e.\ below the ZnTe band gap. In the former case, the excitation source was a solid state laser and in the latter a frequency doubled output of an optical parametric oscillator pumped with a 2 ps Ti:sapphire laser. Single dots were accessed through 200 nm shadow mask apertures produced on the sample surface by spin casting of polystyrene beads and evaporating a 100 nm layer of gold. The laser beam was focused with a microscope objective. The PL signal was collected with the same objective and directed onto a slit of a monochromator. Time-integrated spectra were measured with a CCD camera, while time-resolved PL was detected with a streak camera with overall temporal resolution of 10 ps. All measurements were performed at a temperature of about 10 K.

\begin{figure}[h!]
\includegraphics[angle=-90,width=0.5\textwidth]{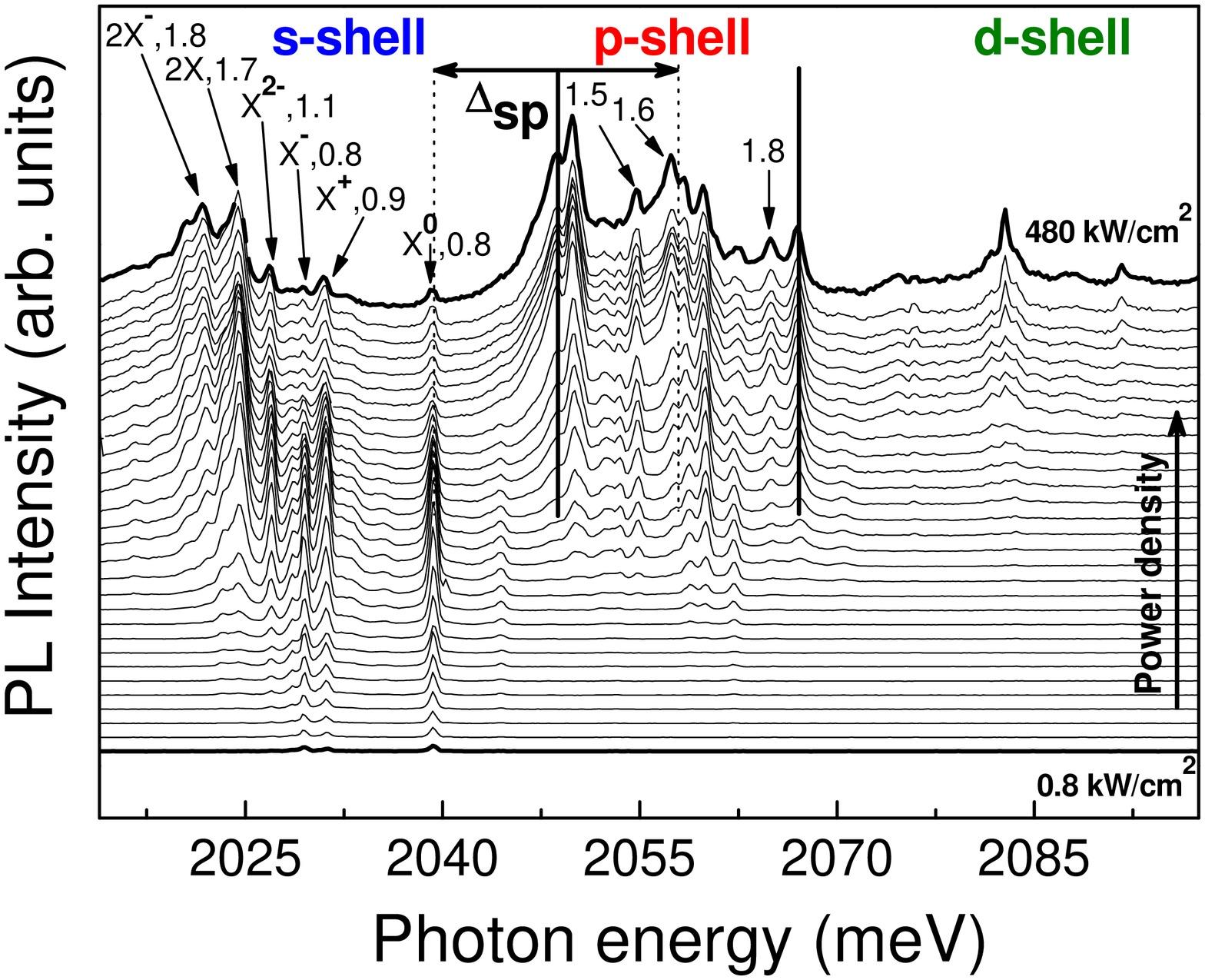}
\caption{PL spectra of a single CdTe quantum dot as a function of cw excitation power density from 0.8 $\rm kW/cm^2$ (bottom spectrum) to 480 $\rm kW/cm^2$ (top spectrum). Exponents of the fitted power law functions are given next to the identified transitions. The method for determining the inter-shell splitting is shown schematically.}
\label{PL_spectra}
\vspace{-1cm}
\end{figure}

In Figure \ref{PL_spectra}, we present cw PL spectra of a single CdTe QD measured with geometrically increasing excitation power density $P$ from 0.8 $\rm kW/cm^2$ (bottom spectrum) to 480 $\rm kW/cm^2$ (top spectrum). For $P \lesssim 30$ $\rm kW/cm^2$ only one emission band, related to the s--shell emission, is observed. At lowest excitation powers, this band consists of three transitions originating from recombinations of the neutral exciton (X$^0$) and charged excitons: \Xp\ and \Xm, which appear red-shifted by 8.2 and 9.7 meV respectively. As the excitation power is increased, the doubly charged exciton X$^{2-}$, the biexciton (2X) and the negatively charged biexciton 2X$^-$ transitions appear, red-shifted with respect to X$^0$ by 12.3, 14.7, and 17.6 meV, respectively. The assignment of these transitions is based on comparison of their respective spectroscopic shifts with those reported for other single CdTe QDs \cite{suf06,leg06,kow06,kaz10,klo11,kaz11}, where the identification was based on photon correlation measurements \cite{suf06,kaz11}, optical anisotropy \cite{kow06,kaz10}, optical orientation \cite{kaz10}, and charging behavior in electric field \cite{leg06,klo11}. It was found that the transition sequence in all QDs with identified charge states is the same, namely $E_{X^0} > E_{X^+} > E_{X^-} > E_{X^{2-}} > E_{2X} > E_{2X^-}$, where $E_{\chi}$ is the emission energy of the complex $\chi$.  The universality of the transition sequence is due to a relatively weak confinement, especially in the valence band, which enhances the role of Coulomb correlation effects. As a result, transitions related to charged excitonic complexes and biexcitons are shifted to lower energies. \cite{klo11} Analogous behavior was observed in weakly confining GaAs QDs made of quantum well width fluctuations \cite{bra05} and in large GaAs dots in \linebreak AlGaAs barriers. \cite{wan09} 

To support our identification of the PL transitions, we fit power law functions to the dependence of their intensities on excitation density. The fitted exponents are given in Fig.\ \ref{PL_spectra} next to the respective transitions. For X$^0$, \Xp, and \Xm\ the PL intensity increases approximately linearly, while for X$^{2-}$, 2X and 2X$^-$ the increase is clearly superlinear in agreement with a typical behavior expected for a neutral exciton and biexciton \cite{bru94,gru97,dek00,suf06,kaz10}.

\begin{figure}[b]
\includegraphics[width=0.5\textwidth]{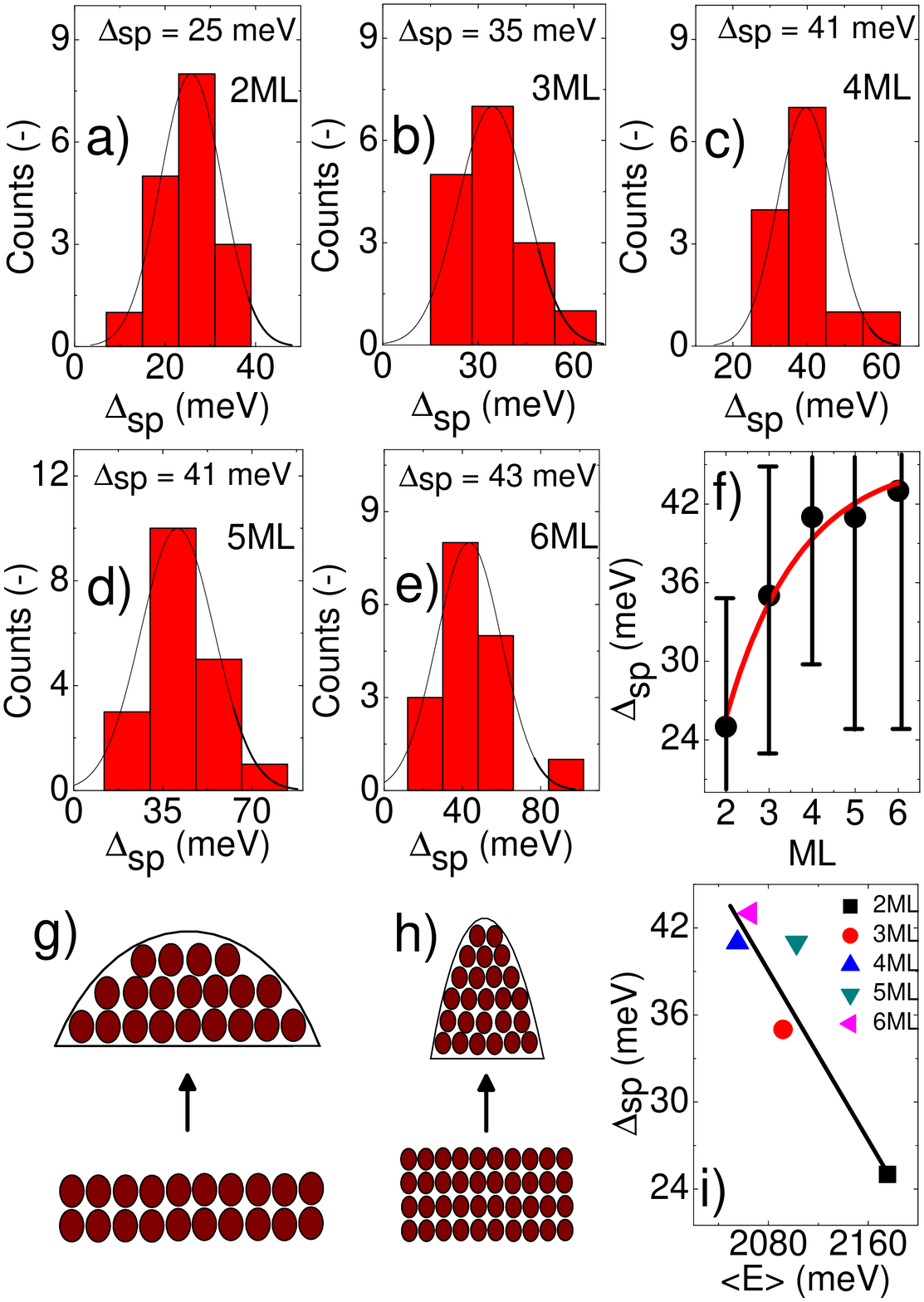}
\caption{(a--e) Distributions of the s--p splitting for different single CdTe QDs for each sample. (f) Evaluated average value of the s--p splitting as a function of the CdTe layer thickness. The error bars indicate the FWHM of the s--p splitting distribution.  (g-h) Formation schemes for QDs with a large and small (small and large) lateral (vertical) size. (i) s-p splitting plotted as a function of the average ensemble emission energy. The solid line is a guide for the eye.}
\label{histograms}
\end{figure}

As the excitation power is increased, at $P \gtrsim 30$ $\rm kW/cm^2$ a higher emission band appears (about 19 meV above X) and simultaneously the neutral and charged excitons begin to lose strength. This band consists of a multitude of transitions, most of which exhibit superlinear intensity increase with excitation power. Thus, we identify it as recombination of p--shell electrons with p--shell holes. Detailed assignment of various transitions to particular QD occupancies require e.g.\ photon correlation measurements and is beyond the scope of this paper. Simultaneously to the appearance of the p--shell emission, a broad low energy tail develops below the 2X$^-$ transition from the s--shell. We attribute this tail to recombinations of s-shell electron and hole in the presence of an occupied p--shell -- an effect also demonstrated in single InGaAs QDs occupied with more than two excitons \cite{bay00,bab06}. With further increase of the excitation density, at about $P \gtrsim 220 $ $\rm kW/cm^2$, another emission band appears, separated from the center of gravity of the p-shell band by about 20 meV. We identify this band with recombination of d--shell electrons with d--shell holes. Since the separation between emission bands related to subsequent shells is approximately equal, we conclude that the shape of the confining lateral potential is approximately parabolic.

In order to determine the impact of the number of CdTe layers on the inter-shell splitting $\Delta_{sp}$, we evaluate the energy distance between the (somewhat arbitrarily chosen -- see Fig.\ \ref{PL_spectra}) center of gravity of the p--shell emission and the X$^0$. Distributions of the obtained $\Delta_{sp}$ values for the five samples are shown in Fig.\ \ref{histograms}a-e together with fitted normal distribution functions. Mean $\Delta_{sp}$ values with errors taken as full width half maxima (FWHM) of the fitted distributions are presented in Fig.\ \ref{histograms}f. A monotonic increase of the inter-shell splitting with the number of CdTe layers from which the QDs are formed is clearly visible.

This increase of $\Delta_{sp}$ points out that with increasing the CdTe layer thickness the lateral size of a dot is decreased. We therefore understand the mechanism of tuning of $\Delta_{sp}$ as the following. With deposition of the strained CdTe layers, the amount of elastic energy accumulated in the sample increases. In a Stranski-Krastanov growth mode, the formation of QDs occurs when this energy is relaxed at the expense of surface energy pertaining to the dots. Therefore, the larger is the amount of elastic energy to be relaxed, the larger can be the resulting QD surface energy and thus the smaller the QD base diameter. Furthermore, with increasing the CdTe layer width we expect the dot height to increase. The formation of dots with large and small base diameters with small and large heights, respectively, is depicted schematically in Figs. \ref{histograms}g-h. The proposed mechanism can thus be confirmed by observing an anticorrelation between the lateral and vertical QD dimensions or conversely an anticorrelation between $\Delta_{sp}$ and mean PL energy, which is governed mostly by the vertical confinement. Indeed, this anticorrelation is observed as displayed in Fig.\ \ref{histograms}i, where mean $\Delta_{sp}$ is plotted against mean emission energy of the QD ensemble.

\begin{figure}[!t]
\includegraphics[angle=-90,width=0.5\textwidth]{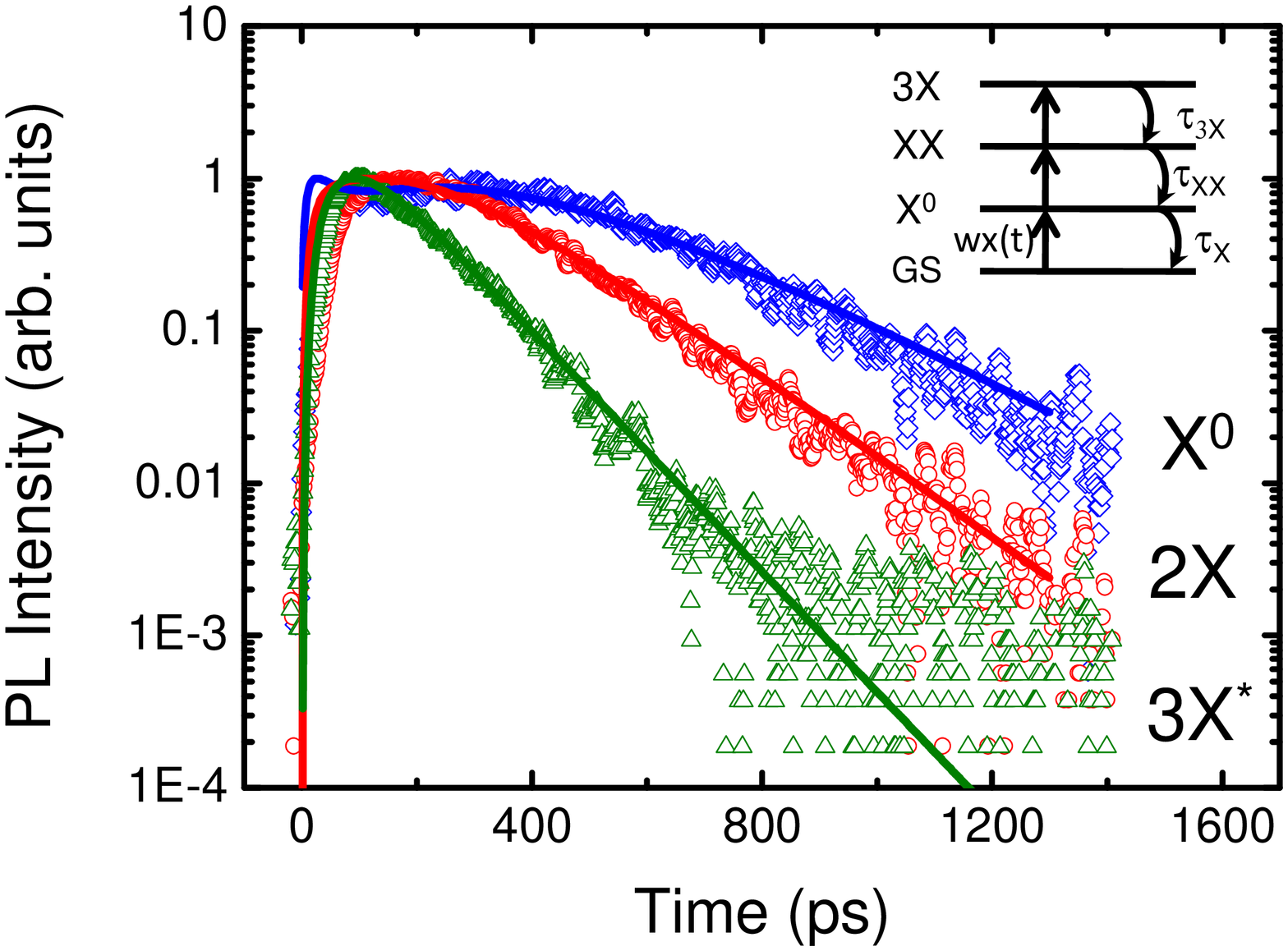}
\caption{Temporal PL traces of the X (diamonds), 2X (circles) and 3X* (triangles) transitions excited with P = 0.8 $\rm kW/cm^2$. Solid lines show the results of rate equation model calculations. The excitation profile is $w_X(t) = G/\tau_{exc} \exp(-t/\tau_{exc}) \theta(t)$, where $G$ is the excitation amplitude, $\theta(t)$ is the Heavside function and the fitted times are: $\tau_X$ = 210 ps, $\tau_{2X}$ = 160 ps, $\tau_{3X^*}$ = 110 ps and $\tau_{exc}$ = 45 ps.}
\label{tr}
\vspace{-0.5cm}
\end{figure}

As seen in Fig.\ \ref{PL_spectra}, emergence of the p--shell emission is accompanied with quenching of the X$^0$. Indeed, under high excitation conditions, the probability of finding the dot occupied by a single electron-hole pair is very small. However, in a time-resolved measurements, it is possible to observe a gradual emptying of the QD of photocarriers. This effect is presented in Fig.\ \ref{tr}, where normalized PL transients related to recombinations of X$^0$, 2X, and a transition from the p--shell marked as 3X$^*$ are shown. The PL rise times are comparable since we employ a non-resonant excitation and a stochastic capture of electrons and holes may result in initially varying QD occupation. On the other hand, the decay dynamics is very different. The 3X$^*$ transition decays monoexponentially, while the 2X and X$^0$ display a plateau due to saturation of the s-shell emission. The plateau is broader for the X$^0$ decay. This behavior is a clear manifestation of a radiative cascade between multiexcitonic states.\cite{san02} Emission of the 3X$^*$ dominates at small delays. Then, when the p-shells are emptied, the 2X takes over, and at longest delays, when the dot is occupied by only one exciton, the X$^0$ dominates.
We model this radiative cascade process by considering X$^0$, 2X, and a neutral 3X transition in a rate equation model \cite{dek00,suf06} for a four level system depicted in the inset to Fig.\ \ref{tr}. The excitation profile is given by an exponential decay with a characteristic exciton capture time \cite{suf06}. Thus, there are five fitting parameters: the excitation amplitude, the three decay times and the exciton capture time. The results are presented as solid lines and fitted times are given in the caption. The ratio between the 2X and X$^0$ lifetimes is about 0.76 demonstrating a strong influence of Coulomb interactions on excitonic wave functions.\cite{klo11} The observed cascaded emission process proves that the CdTe QDs can be utilized as sources of non-classical light. \cite{suf06}

To conclude, we studied emission properties of highly excited single CdTe QDs. We observed sequential filling of the shells accompanied by appearance of higher emission bands. We evaluated the splitting between the s- and p-shell emission for about 100 QDs formed from CdTe layers of different thicknesses and demonstrated a new tool for tuning the inter-shell splitting in CdTe QDs. Namely, we found that with increasing CdTe layer thickness, the s-p splitting increases. Moreover, in a time-resolved PL measurement, we observed a radiative cascade between multiexcitonic complexes, in which the filling of the s-shell led to a saturation of the X$^0$ and 2X transitions and their subsequent emergence at higher delays. The result were accounted for in a rate equation model.

This work was supported by European Regional Development Fund through grant Innovative Economy (POIG.01.01.02-00-008/08) and by the Polonium Programme.


\begin{thebibliography}{26}%
\makeatletter
\providecommand \@ifxundefined [1]{%
 \@ifx{#1\undefined}
}%
\providecommand \@ifnum [1]{%
 \ifnum #1\expandafter \@firstoftwo
 \else \expandafter \@secondoftwo
 \fi
}%
\providecommand \@ifx [1]{%
 \ifx #1\expandafter \@firstoftwo
 \else \expandafter \@secondoftwo
 \fi
}%
\providecommand \natexlab [1]{#1}%
\providecommand \enquote  [1]{``#1''}%
\providecommand \bibnamefont  [1]{#1}%
\providecommand \bibfnamefont [1]{#1}%
\providecommand \citenamefont [1]{#1}%
\providecommand \href@noop [0]{\@secondoftwo}%
\providecommand \href [0]{\begingroup \@sanitize@url \@href}%
\providecommand \@href[1]{\@@startlink{#1}\@@href}%
\providecommand \@@href[1]{\endgroup#1\@@endlink}%
\providecommand \@sanitize@url [0]{\catcode `\\12\catcode `\$12\catcode
  `\&12\catcode `\#12\catcode `\^12\catcode `\_12\catcode `\%12\relax}%
\providecommand \@@startlink[1]{}%
\providecommand \@@endlink[0]{}%
\providecommand \url  [0]{\begingroup\@sanitize@url \@url }%
\providecommand \@url [1]{\endgroup\@href {#1}{\urlprefix }}%
\providecommand \urlprefix  [0]{URL }%
\providecommand \Eprint [0]{\href }%
\@ifxundefined \urlstyle {%
  \providecommand \doi  [0]{\begingroup \@sanitize@url \@doi}%
  \providecommand \@doi [1]{\endgroup \@@startlink {\doibase
  #1}doi:\discretionary {}{}{}#1\@@endlink }%
}{%
  \providecommand \doi  [0]{doi:\discretionary{}{}{}\begingroup
  \urlstyle{rm}\Url }%
}%
\providecommand \doibase [0]{http://dx.doi.org/}%
\providecommand \Doi [0]{\begingroup \@sanitize@url \@Doi }%
\providecommand \@Doi  [1]{\endgroup\@@startlink{\doibase#1}\@@Doi}%
\providecommand \@@Doi [1]{#1\@@endlink}%
\providecommand \selectlanguage [0]{\@gobble}%
\providecommand \bibinfo  [0]{\@secondoftwo}%
\providecommand \bibfield  [0]{\@secondoftwo}%
\providecommand \translation [1]{[#1]}%
\providecommand \BibitemOpen [0]{}%
\providecommand \bibitemStop [0]{}%
\providecommand \bibitemNoStop [0]{.\EOS\space}%
\providecommand \EOS [0]{\spacefactor3000\relax}%
\providecommand \BibitemShut  [1]{\csname bibitem#1\endcsname}%
\bibitem [{\citenamefont {Jacak}\ \emph {et~al.}(1998)\citenamefont {Jacak},
  \citenamefont {Hawrylak},\ and\ \citenamefont {Wójs}}]{jac98}%
  \BibitemOpen
  \bibfield  {author} {\bibinfo {author} {\bibfnamefont {L.}~\bibnamefont
  {Jacak}}, \bibinfo {author} {\bibfnamefont {P.}~\bibnamefont {Hawrylak}}, \
  and\ \bibinfo {author} {\bibfnamefont {A.}~\bibnamefont {Wójs}},\ }\href@noop
  {} {\emph {\bibinfo {title} {Quantum Dots}}}\ (\bibinfo  {publisher}
  {Springer},\ \bibinfo {year} {1998})\BibitemShut {NoStop}%
\bibitem [{\citenamefont {Michler}(2009)}]{mic09}%
  \BibitemOpen
  \bibinfo {editor} {\bibfnamefont {P.}~\bibnamefont {Michler}},\ ed.,\
  \href@noop {} {\emph {\bibinfo {title} {Single Semiconductor Quantum Dots}}}\
  (\bibinfo  {publisher} {Springer},\ \bibinfo {year} {2009})\BibitemShut
  {NoStop}%
\bibitem [{\citenamefont {Khaetskii}\ and\ \citenamefont
  {Nazarov}(2000)}]{kha00}%
  \BibitemOpen
  \bibfield  {author} {\bibinfo {author} {\bibfnamefont {A.~V.}\ \bibnamefont
  {Khaetskii}}\ and\ \bibinfo {author} {\bibfnamefont {Y.~V.}\ \bibnamefont
  {Nazarov}},\ }\href@noop {} {\bibfield  {journal} {\bibinfo  {journal} {Phys.
  Rev. B},\ }\textbf {\bibinfo {volume} {61}},\ \bibinfo {pages} {12639}
  (\bibinfo {year} {2000})}\BibitemShut {NoStop}%
\bibitem [{\citenamefont {Michler}\ \emph {et~al.}(2000)\citenamefont
  {Michler}, \citenamefont {Kiraz}, \citenamefont {Becher}, \citenamefont
  {Schoenfeld}, \citenamefont {Petroff}, \citenamefont {Zhang}, \citenamefont
  {Hu},\ and\ \citenamefont {Imamo\u{g}lu}}]{mic00}%
  \BibitemOpen
  \bibfield  {author} {\bibinfo {author} {\bibfnamefont {P.}~\bibnamefont
  {Michler}}, \bibinfo {author} {\bibfnamefont {A.}~\bibnamefont {Kiraz}},
  \bibinfo {author} {\bibfnamefont {C.}~\bibnamefont {Becher}}, \bibinfo
  {author} {\bibfnamefont {W.~V.}\ \bibnamefont {Schoenfeld}}, \bibinfo
  {author} {\bibfnamefont {P.~M.}\ \bibnamefont {Petroff}}, \bibinfo {author}
  {\bibfnamefont {L.}~\bibnamefont {Zhang}}, \bibinfo {author} {\bibfnamefont
  {E.}~\bibnamefont {Hu}}, \ and\ \bibinfo {author} {\bibfnamefont
  {A.}~\bibnamefont {Imamo\u{g}lu}},\ }\href@noop {} {\bibfield  {journal}
  {\bibinfo  {journal} {Science},\ }\textbf {\bibinfo {volume} {290}},\
  \bibinfo {pages} {2292} (\bibinfo {year} {2000})}\BibitemShut {NoStop}%
\bibitem [{\citenamefont {Fafard}\ and\ \citenamefont {Allen}(1999)}]{faf99}%
  \BibitemOpen
  \bibfield  {author} {\bibinfo {author} {\bibfnamefont {S.}~\bibnamefont
  {Fafard}}\ and\ \bibinfo {author} {\bibfnamefont {C.~N.}\ \bibnamefont
  {Allen}},\ }\href@noop {} {\bibfield  {journal} {\bibinfo  {journal} {Appl.
  Phys. Lett.},\ }\textbf {\bibinfo {volume} {75}},\ \bibinfo {pages} {2374}
  (\bibinfo {year} {1999})}\BibitemShut {NoStop}%
\bibitem [{\citenamefont {Allen}\ \emph {et~al.}(2001)\citenamefont {Allen},
  \citenamefont {Finnie}, \citenamefont {Raymond}, \citenamefont {Wasilewski},\
  and\ \citenamefont {Fafard}}]{all01}%
  \BibitemOpen
  \bibfield  {author} {\bibinfo {author} {\bibfnamefont {C.~N.}\ \bibnamefont
  {Allen}}, \bibinfo {author} {\bibfnamefont {P.}~\bibnamefont {Finnie}},
  \bibinfo {author} {\bibfnamefont {S.}~\bibnamefont {Raymond}}, \bibinfo
  {author} {\bibfnamefont {Z.~R.}\ \bibnamefont {Wasilewski}}, \ and\ \bibinfo
  {author} {\bibfnamefont {S.}~\bibnamefont {Fafard}},\ }\href@noop {}
  {\bibfield  {journal} {\bibinfo  {journal} {Appl. Phys. Lett.},\ }\textbf
  {\bibinfo {volume} {79}},\ \bibinfo {pages} {2701} (\bibinfo {year}
  {2001})}\BibitemShut {NoStop}%
\bibitem [{\citenamefont {Maćkowski}\ \emph {et~al.}(2003)\citenamefont
  {Maćkowski}, \citenamefont {Smith},\ and\ \citenamefont {Jackson}}]{mac03}%
  \BibitemOpen
  \bibfield  {author} {\bibinfo {author} {\bibfnamefont {S.}~\bibnamefont
  {Maćkowski}}, \bibinfo {author} {\bibfnamefont {L.~M.}\ \bibnamefont
  {Smith}}, \ and\ \bibinfo {author} {\bibfnamefont {H.~E.}\ \bibnamefont
  {Jackson}},\ }\href@noop {} {\bibfield  {journal} {\bibinfo  {journal} {Appl.
  Phys. Lett.},\ }\textbf {\bibinfo {volume} {83}},\ \bibinfo {pages} {254}
  (\bibinfo {year} {2003})}\BibitemShut {NoStop}%
\bibitem [{\citenamefont {Wojnar}\ \emph {et~al.}(2007)\citenamefont {Wojnar},
  \citenamefont {Karczewski}, \citenamefont {Wojtowicz},\ and\ \citenamefont
  {Kossut}}]{woj07}%
  \BibitemOpen
  \bibfield  {author} {\bibinfo {author} {\bibfnamefont {P.}~\bibnamefont
  {Wojnar}}, \bibinfo {author} {\bibfnamefont {G.}~\bibnamefont {Karczewski}},
  \bibinfo {author} {\bibfnamefont {T.}~\bibnamefont {Wojtowicz}}, \ and\
  \bibinfo {author} {\bibfnamefont {J.}~\bibnamefont {Kossut}},\ }\href@noop {}
  {\bibfield  {journal} {\bibinfo  {journal} {Acta Phys. Pol. A},\ }\textbf
  {\bibinfo {volume} {112}},\ \bibinfo {pages} {283} (\bibinfo {year}
  {2007})}\BibitemShut {NoStop}%
\bibitem [{\citenamefont {Fafard}\ \emph
  {et~al.}(1999){\natexlab{a}}\citenamefont {Fafard}, \citenamefont
  {Wasilewski}, \citenamefont {Allen}, \citenamefont {Picard}, \citenamefont
  {Spanner}, \citenamefont {McCaffrey},\ and\ \citenamefont {Piva}}]{faf99prb}%
  \BibitemOpen
  \bibfield  {author} {\bibinfo {author} {\bibfnamefont {S.}~\bibnamefont
  {Fafard}}, \bibinfo {author} {\bibfnamefont {Z.~R.}\ \bibnamefont
  {Wasilewski}}, \bibinfo {author} {\bibfnamefont {C.~N.}\ \bibnamefont
  {Allen}}, \bibinfo {author} {\bibfnamefont {D.}~\bibnamefont {Picard}},
  \bibinfo {author} {\bibfnamefont {M.}~\bibnamefont {Spanner}}, \bibinfo
  {author} {\bibfnamefont {J.~P.}\ \bibnamefont {McCaffrey}}, \ and\ \bibinfo
  {author} {\bibfnamefont {P.~G.}\ \bibnamefont {Piva}},\ }\Doi
  {10.1103/PhysRevB.59.15368} {\bibfield  {journal} {\bibinfo  {journal} {Phys.
  Rev. B},\ }\textbf {\bibinfo {volume} {59}},\ \bibinfo {pages} {15368}
  (\bibinfo {year} {1999}{\natexlab{a}})}\BibitemShut {NoStop}%
\bibitem [{\citenamefont {Wasilewski}\ \emph {et~al.}(1999)\citenamefont
  {Wasilewski}, \citenamefont {Fafard},\ and\ \citenamefont
  {McCaffrey}}]{was99}%
  \BibitemOpen
  \bibfield  {author} {\bibinfo {author} {\bibfnamefont {Z.}~\bibnamefont
  {Wasilewski}}, \bibinfo {author} {\bibfnamefont {S.}~\bibnamefont {Fafard}},
  \ and\ \bibinfo {author} {\bibfnamefont {J.}~\bibnamefont {McCaffrey}},\
  }\href@noop {} {\bibfield  {journal} {\bibinfo  {journal} {J. Cryst.
  Growth},\ }\textbf {\bibinfo {volume} {201/202}},\ \bibinfo {pages} {1131}
  (\bibinfo {year} {1999})}\BibitemShut {NoStop}%
\bibitem [{\citenamefont {Fafard}\ \emph
  {et~al.}(1999){\natexlab{b}}\citenamefont {Fafard}, \citenamefont
  {Wasilewski}, \citenamefont {Allen}, \citenamefont {Hinzer}, \citenamefont
  {McCaffrey},\ and\ \citenamefont {Feng}}]{faf99lasing}%
  \BibitemOpen
  \bibfield  {author} {\bibinfo {author} {\bibfnamefont {S.}~\bibnamefont
  {Fafard}}, \bibinfo {author} {\bibfnamefont {Z.~R.}\ \bibnamefont
  {Wasilewski}}, \bibinfo {author} {\bibfnamefont {C.~N.}\ \bibnamefont
  {Allen}}, \bibinfo {author} {\bibfnamefont {K.}~\bibnamefont {Hinzer}},
  \bibinfo {author} {\bibfnamefont {J.~P.}\ \bibnamefont {McCaffrey}}, \ and\
  \bibinfo {author} {\bibfnamefont {Y.}~\bibnamefont {Feng}},\ }\href@noop {}
  {\textbf {\bibinfo {volume} {75}},\ \bibinfo {pages} {986} (\bibinfo {year}
  {1999}{\natexlab{b}})}\BibitemShut {NoStop}%
\bibitem [{\citenamefont {Wang}\ \emph {et~al.}(2009)\citenamefont {Wang},
  \citenamefont {K\ifmmode~\check{r}\else \v{r}\fi{}\'apek}, \citenamefont
  {Ding}, \citenamefont {Horton}, \citenamefont {Schliwa}, \citenamefont
  {Bimberg}, \citenamefont {Rastelli},\ and\ \citenamefont {Schmidt}}]{wan09}%
  \BibitemOpen
  \bibfield  {author} {\bibinfo {author} {\bibfnamefont {L.}~\bibnamefont
  {Wang}}, \bibinfo {author} {\bibfnamefont {V.}~\bibnamefont
  {K\ifmmode~\check{r}\else \v{r}\fi{}\'apek}}, \bibinfo {author}
  {\bibfnamefont {F.}~\bibnamefont {Ding}}, \bibinfo {author} {\bibfnamefont
  {F.}~\bibnamefont {Horton}}, \bibinfo {author} {\bibfnamefont
  {A.}~\bibnamefont {Schliwa}}, \bibinfo {author} {\bibfnamefont
  {D.}~\bibnamefont {Bimberg}}, \bibinfo {author} {\bibfnamefont
  {A.}~\bibnamefont {Rastelli}}, \ and\ \bibinfo {author} {\bibfnamefont
  {O.~G.}\ \bibnamefont {Schmidt}},\ }\Doi {10.1103/PhysRevB.80.085309}
  {\bibfield  {journal} {\bibinfo  {journal} {Phys. Rev. B},\ }\textbf
  {\bibinfo {volume} {80}},\ \bibinfo {pages} {085309} (\bibinfo {year}
  {2009})}\BibitemShut {NoStop}%
\bibitem [{\citenamefont {Tinjod}\ \emph {et~al.}(2003)\citenamefont {Tinjod},
  \citenamefont {Gilles}, \citenamefont {Moehl}, \citenamefont {Kheng},\ and\
  \citenamefont {Mariette}}]{tin03}%
  \BibitemOpen
  \bibfield  {author} {\bibinfo {author} {\bibfnamefont {F.}~\bibnamefont
  {Tinjod}}, \bibinfo {author} {\bibfnamefont {B.}~\bibnamefont {Gilles}},
  \bibinfo {author} {\bibfnamefont {S.}~\bibnamefont {Moehl}}, \bibinfo
  {author} {\bibfnamefont {K.}~\bibnamefont {Kheng}}, \ and\ \bibinfo {author}
  {\bibfnamefont {H.}~\bibnamefont {Mariette}},\ }\href@noop {} {\bibfield
  {journal} {\bibinfo  {journal} {Appl. Phys. Lett.},\ }\textbf {\bibinfo
  {volume} {82}},\ \bibinfo {pages} {4340} (\bibinfo {year}
  {2003})}\BibitemShut {NoStop}%
\bibitem [{\citenamefont {Suffczyński}\ \emph {et~al.}(2006)\citenamefont
  {Suffczyński}, \citenamefont {Kazimierczuk}, \citenamefont {Goryca},
  \citenamefont {Piechal}, \citenamefont {Trajnerowicz}, \citenamefont
  {Kowalik}, \citenamefont {Kossacki}, \citenamefont {Golnik}, \citenamefont
  {Korona}, \citenamefont {Nawrocki}, \citenamefont {Gaj},\ and\ \citenamefont
  {Karczewski}}]{suf06}%
  \BibitemOpen
  \bibfield  {author} {\bibinfo {author} {\bibfnamefont {J.}~\bibnamefont
  {Suffczyński}}, \bibinfo {author} {\bibfnamefont {T.}~\bibnamefont
  {Kazimierczuk}}, \bibinfo {author} {\bibfnamefont {M.}~\bibnamefont
  {Goryca}}, \bibinfo {author} {\bibfnamefont {B.}~\bibnamefont {Piechal}},
  \bibinfo {author} {\bibfnamefont {A.}~\bibnamefont {Trajnerowicz}}, \bibinfo
  {author} {\bibfnamefont {K.}~\bibnamefont {Kowalik}}, \bibinfo {author}
  {\bibfnamefont {P.}~\bibnamefont {Kossacki}}, \bibinfo {author}
  {\bibfnamefont {A.}~\bibnamefont {Golnik}}, \bibinfo {author} {\bibfnamefont
  {K.~P.}\ \bibnamefont {Korona}}, \bibinfo {author} {\bibfnamefont
  {M.}~\bibnamefont {Nawrocki}}, \bibinfo {author} {\bibfnamefont {J.~A.}\
  \bibnamefont {Gaj}}, \ and\ \bibinfo {author} {\bibfnamefont
  {G.}~\bibnamefont {Karczewski}},\ }\Doi {10.1103/PhysRevB.74.085319}
  {\bibfield  {journal} {\bibinfo  {journal} {Phys. Rev. B},\ }\textbf
  {\bibinfo {volume} {74}},\ \bibinfo {pages} {085319} (\bibinfo {year}
  {2006})}\BibitemShut {NoStop}%
\bibitem [{\citenamefont {L\'eger}\ \emph {et~al.}(2006)\citenamefont
  {L\'eger}, \citenamefont {Besombes}, \citenamefont {Fern\'andez-Rossier},
  \citenamefont {Maingault},\ and\ \citenamefont {Mariette}}]{leg06}%
  \BibitemOpen
  \bibfield  {author} {\bibinfo {author} {\bibfnamefont {Y.}~\bibnamefont
  {L\'eger}}, \bibinfo {author} {\bibfnamefont {L.}~\bibnamefont {Besombes}},
  \bibinfo {author} {\bibfnamefont {J.}~\bibnamefont {Fern\'andez-Rossier}},
  \bibinfo {author} {\bibfnamefont {L.}~\bibnamefont {Maingault}}, \ and\
  \bibinfo {author} {\bibfnamefont {H.}~\bibnamefont {Mariette}},\ }\Doi
  {10.1103/PhysRevLett.97.107401} {\bibfield  {journal} {\bibinfo  {journal}
  {Phys. Rev. Lett.},\ }\textbf {\bibinfo {volume} {97}},\ \bibinfo {pages}
  {107401} (\bibinfo {year} {2006})}\BibitemShut {NoStop}%
\bibitem [{\citenamefont {Kowalik}\ \emph {et~al.}(2006)\citenamefont
  {Kowalik}, \citenamefont {Kudelski}, \citenamefont {Golnik}, \citenamefont
  {Suffczyński}, \citenamefont {Krebs}, \citenamefont {Voisin}, \citenamefont
  {Karczewski}, \citenamefont {Kossut},\ and\ \citenamefont {Gaj}}]{kow06}%
  \BibitemOpen
  \bibfield  {author} {\bibinfo {author} {\bibfnamefont {K.}~\bibnamefont
  {Kowalik}}, \bibinfo {author} {\bibfnamefont {A.}~\bibnamefont {Kudelski}},
  \bibinfo {author} {\bibfnamefont {A.}~\bibnamefont {Golnik}}, \bibinfo
  {author} {\bibfnamefont {J.}~\bibnamefont {Suffczyński}}, \bibinfo {author}
  {\bibfnamefont {O.}~\bibnamefont {Krebs}}, \bibinfo {author} {\bibfnamefont
  {P.}~\bibnamefont {Voisin}}, \bibinfo {author} {\bibfnamefont
  {G.}~\bibnamefont {Karczewski}}, \bibinfo {author} {\bibfnamefont
  {J.}~\bibnamefont {Kossut}}, \ and\ \bibinfo {author} {\bibfnamefont
  {J.}~\bibnamefont {Gaj}},\ }\href@noop {} {\bibfield  {journal} {\bibinfo
  {journal} {phys. stat. sol. c},\ }\textbf {\bibinfo {volume} {3}},\ \bibinfo
  {pages} {865} (\bibinfo {year} {2006})}\BibitemShut {NoStop}%
\bibitem [{\citenamefont {Kazimierczuk}\ \emph {et~al.}(2010)\citenamefont
  {Kazimierczuk}, \citenamefont {Goryca}, \citenamefont {Koperski},
  \citenamefont {Golnik}, \citenamefont {Gaj}, \citenamefont {Nawrocki},
  \citenamefont {Wojnar},\ and\ \citenamefont {Kossacki}}]{kaz10}%
  \BibitemOpen
  \bibfield  {author} {\bibinfo {author} {\bibfnamefont {T.}~\bibnamefont
  {Kazimierczuk}}, \bibinfo {author} {\bibfnamefont {M.}~\bibnamefont
  {Goryca}}, \bibinfo {author} {\bibfnamefont {M.}~\bibnamefont {Koperski}},
  \bibinfo {author} {\bibfnamefont {A.}~\bibnamefont {Golnik}}, \bibinfo
  {author} {\bibfnamefont {J.~A.}\ \bibnamefont {Gaj}}, \bibinfo {author}
  {\bibfnamefont {M.}~\bibnamefont {Nawrocki}}, \bibinfo {author}
  {\bibfnamefont {P.}~\bibnamefont {Wojnar}}, \ and\ \bibinfo {author}
  {\bibfnamefont {P.}~\bibnamefont {Kossacki}},\ }\Doi
  {10.1103/PhysRevB.81.155313} {\bibfield  {journal} {\bibinfo  {journal}
  {Phys. Rev. B},\ }\textbf {\bibinfo {volume} {81}},\ \bibinfo {pages}
  {155313} (\bibinfo {year} {2010})}\BibitemShut {NoStop}%
\bibitem [{\citenamefont {K\l{}opotowski}\ \emph {et~al.}(2011)\citenamefont
  {K\l{}opotowski}, \citenamefont {Voliotis}, \citenamefont {Kudelski},
  \citenamefont {Tartakovskii}, \citenamefont {Wojnar}, \citenamefont {Fronc},
  \citenamefont {Grousson}, \citenamefont {Krebs}, \citenamefont {Skolnick},
  \citenamefont {Karczewski},\ and\ \citenamefont {Wojtowicz}}]{klo11}%
  \BibitemOpen
  \bibfield  {author} {\bibinfo {author} {\bibfnamefont {{\L}.}~\bibnamefont
  {K\l{}opotowski}}, \bibinfo {author} {\bibfnamefont {V.}~\bibnamefont
  {Voliotis}}, \bibinfo {author} {\bibfnamefont {A.}~\bibnamefont {Kudelski}},
  \bibinfo {author} {\bibfnamefont {A.~I.}\ \bibnamefont {Tartakovskii}},
  \bibinfo {author} {\bibfnamefont {P.}~\bibnamefont {Wojnar}}, \bibinfo
  {author} {\bibfnamefont {K.}~\bibnamefont {Fronc}}, \bibinfo {author}
  {\bibfnamefont {R.}~\bibnamefont {Grousson}}, \bibinfo {author}
  {\bibfnamefont {O.}~\bibnamefont {Krebs}}, \bibinfo {author} {\bibfnamefont
  {M.~S.}\ \bibnamefont {Skolnick}}, \bibinfo {author} {\bibfnamefont
  {G.}~\bibnamefont {Karczewski}}, \ and\ \bibinfo {author} {\bibfnamefont
  {T.}~\bibnamefont {Wojtowicz}},\ }\href@noop {} {\bibfield  {journal}
  {\bibinfo  {journal} {Phys. Rev. B},\ }\textbf {\bibinfo {volume} {83}},\
  \bibinfo {pages} {155319} (\bibinfo {year} {2011})}\BibitemShut {NoStop}%
\bibitem [{\citenamefont {Kazimierczuk}\ \emph {et~al.}(2011)\citenamefont
  {Kazimierczuk}, \citenamefont {Smolenski}, \citenamefont {Goryca},
  \citenamefont {K{\l}opotowski}, \citenamefont {Wojnar}, \citenamefont
  {Fronc}, \citenamefont {Golnik}, \citenamefont {Nawrocki}, \citenamefont
  {Gaj},\ and\ \citenamefont {Kossacki}}]{kaz11}%
  \BibitemOpen
  \bibfield  {author} {\bibinfo {author} {\bibfnamefont {T.}~\bibnamefont
  {Kazimierczuk}}, \bibinfo {author} {\bibfnamefont {T.}~\bibnamefont
  {Smolenski}}, \bibinfo {author} {\bibfnamefont {M.}~\bibnamefont {Goryca}},
  \bibinfo {author} {\bibfnamefont {{\L}.}~\bibnamefont {K{\l}opotowski}},
  \bibinfo {author} {\bibfnamefont {P.}~\bibnamefont {Wojnar}}, \bibinfo
  {author} {\bibfnamefont {K.}~\bibnamefont {Fronc}}, \bibinfo {author}
  {\bibfnamefont {A.}~\bibnamefont {Golnik}}, \bibinfo {author} {\bibfnamefont
  {M.}~\bibnamefont {Nawrocki}}, \bibinfo {author} {\bibfnamefont
  {J.}~\bibnamefont {Gaj}}, \ and\ \bibinfo {author} {\bibfnamefont
  {P.}~\bibnamefont {Kossacki}},\ }\href@noop {} {\bibfield  {journal}
  {\bibinfo  {journal} {Phys. Rev. B},\ \bibinfo {pages} {submitted}} (\bibinfo
  {year} {2011})}\BibitemShut {NoStop}%
\bibitem [{\citenamefont {Bracker}\ \emph {et~al.}(2005)\citenamefont
  {Bracker}, \citenamefont {Stinaff}, \citenamefont {Gammon}, \citenamefont
  {Ware}, \citenamefont {Tischler}, \citenamefont {Park}, \citenamefont
  {Gershoni}, \citenamefont {Filinov}, \citenamefont {Bonitz}, \citenamefont
  {Peeters},\ and\ \citenamefont {Riva}}]{bra05}%
  \BibitemOpen
  \bibfield  {author} {\bibinfo {author} {\bibfnamefont {A.~S.}\ \bibnamefont
  {Bracker}}, \bibinfo {author} {\bibfnamefont {E.~A.}\ \bibnamefont
  {Stinaff}}, \bibinfo {author} {\bibfnamefont {D.}~\bibnamefont {Gammon}},
  \bibinfo {author} {\bibfnamefont {M.~E.}\ \bibnamefont {Ware}}, \bibinfo
  {author} {\bibfnamefont {J.~G.}\ \bibnamefont {Tischler}}, \bibinfo {author}
  {\bibfnamefont {D.}~\bibnamefont {Park}}, \bibinfo {author} {\bibfnamefont
  {D.}~\bibnamefont {Gershoni}}, \bibinfo {author} {\bibfnamefont {A.~V.}\
  \bibnamefont {Filinov}}, \bibinfo {author} {\bibfnamefont {M.}~\bibnamefont
  {Bonitz}}, \bibinfo {author} {\bibfnamefont {F.}~\bibnamefont {Peeters}}, \
  and\ \bibinfo {author} {\bibfnamefont {C.}~\bibnamefont {Riva}},\ }\Doi
  {10.1103/PhysRevB.72.035332} {\bibfield  {journal} {\bibinfo  {journal}
  {Phys. Rev. B},\ }\textbf {\bibinfo {volume} {72}},\ \bibinfo {pages}
  {035332} (\bibinfo {year} {2005})}\BibitemShut {NoStop}%
\bibitem [{\citenamefont {Brunner}\ \emph {et~al.}(1994)\citenamefont
  {Brunner}, \citenamefont {Abstreiter}, \citenamefont {B\"ohm}, \citenamefont
  {Tr\"ankle},\ and\ \citenamefont {Weimann}}]{bru94}%
  \BibitemOpen
  \bibfield  {author} {\bibinfo {author} {\bibfnamefont {K.}~\bibnamefont
  {Brunner}}, \bibinfo {author} {\bibfnamefont {G.}~\bibnamefont {Abstreiter}},
  \bibinfo {author} {\bibfnamefont {G.}~\bibnamefont {B\"ohm}}, \bibinfo
  {author} {\bibfnamefont {G.}~\bibnamefont {Tr\"ankle}}, \ and\ \bibinfo
  {author} {\bibfnamefont {G.}~\bibnamefont {Weimann}},\ }\Doi
  {10.1103/PhysRevLett.73.1138} {\bibfield  {journal} {\bibinfo  {journal}
  {Phys. Rev. Lett.},\ }\textbf {\bibinfo {volume} {73}},\ \bibinfo {pages}
  {1138} (\bibinfo {year} {1994})}\BibitemShut {NoStop}%
\bibitem [{\citenamefont {Grundmann}\ and\ \citenamefont
  {Bimberg}(1997)}]{gru97}%
  \BibitemOpen
  \bibfield  {author} {\bibinfo {author} {\bibfnamefont {M.}~\bibnamefont
  {Grundmann}}\ and\ \bibinfo {author} {\bibfnamefont {D.}~\bibnamefont
  {Bimberg}},\ }\Doi {10.1103/PhysRevB.55.9740} {\bibfield  {journal} {\bibinfo
   {journal} {Phys. Rev. B},\ }\textbf {\bibinfo {volume} {55}},\ \bibinfo
  {pages} {9740} (\bibinfo {year} {1997})}\BibitemShut {NoStop}%
\bibitem [{\citenamefont {Dekel}\ \emph {et~al.}(2000)\citenamefont {Dekel},
  \citenamefont {Regelman}, \citenamefont {Gershoni}, \citenamefont
  {Ehrenfreund}, \citenamefont {Schoenfeld},\ and\ \citenamefont
  {Petroff}}]{dek00}%
  \BibitemOpen
  \bibfield  {author} {\bibinfo {author} {\bibfnamefont {E.}~\bibnamefont
  {Dekel}}, \bibinfo {author} {\bibfnamefont {D.~V.}\ \bibnamefont {Regelman}},
  \bibinfo {author} {\bibfnamefont {D.}~\bibnamefont {Gershoni}}, \bibinfo
  {author} {\bibfnamefont {E.}~\bibnamefont {Ehrenfreund}}, \bibinfo {author}
  {\bibfnamefont {W.~V.}\ \bibnamefont {Schoenfeld}}, \ and\ \bibinfo {author}
  {\bibfnamefont {P.~M.}\ \bibnamefont {Petroff}},\ }\Doi
  {10.1103/PhysRevB.62.11038} {\bibfield  {journal} {\bibinfo  {journal} {Phys.
  Rev. B},\ }\textbf {\bibinfo {volume} {62}},\ \bibinfo {pages} {11038}
  (\bibinfo {year} {2000})}\BibitemShut {NoStop}%
\bibitem [{\citenamefont {Bayer}\ \emph {et~al.}(2000)\citenamefont {Bayer},
  \citenamefont {Stern}, \citenamefont {Hawrylak}, \citenamefont {Fafard},\
  and\ \citenamefont {Forchel}}]{bay00}%
  \BibitemOpen
  \bibfield  {author} {\bibinfo {author} {\bibfnamefont {M.}~\bibnamefont
  {Bayer}}, \bibinfo {author} {\bibfnamefont {O.}~\bibnamefont {Stern}},
  \bibinfo {author} {\bibfnamefont {P.}~\bibnamefont {Hawrylak}}, \bibinfo
  {author} {\bibfnamefont {S.}~\bibnamefont {Fafard}}, \ and\ \bibinfo {author}
  {\bibfnamefont {A.}~\bibnamefont {Forchel}},\ }\href@noop {} {\bibfield
  {journal} {\bibinfo  {journal} {Nature},\ }\textbf {\bibinfo {volume}
  {405}},\ \bibinfo {pages} {923} (\bibinfo {year} {2000})}\BibitemShut
  {NoStop}%
\bibitem [{\citenamefont {Babiński}\ \emph {et~al.}(2006)\citenamefont
  {Babiński}, \citenamefont {Potemski}, \citenamefont {Raymond}, \citenamefont
  {Lapointe},\ and\ \citenamefont {Wasilewski}}]{bab06}%
  \BibitemOpen
  \bibfield  {author} {\bibinfo {author} {\bibfnamefont {A.}~\bibnamefont
  {Babiński}}, \bibinfo {author} {\bibfnamefont {M.}~\bibnamefont {Potemski}},
  \bibinfo {author} {\bibfnamefont {S.}~\bibnamefont {Raymond}}, \bibinfo
  {author} {\bibfnamefont {J.}~\bibnamefont {Lapointe}}, \ and\ \bibinfo
  {author} {\bibfnamefont {Z.~R.}\ \bibnamefont {Wasilewski}},\ }\Doi
  {10.1103/PhysRevB.74.155301} {\bibfield  {journal} {\bibinfo  {journal}
  {Phys. Rev. B},\ }\textbf {\bibinfo {volume} {74}},\ \bibinfo {pages}
  {155301} (\bibinfo {year} {2006})}\BibitemShut {NoStop}%
\bibitem [{\citenamefont {Santori}\ \emph {et~al.}(2002)\citenamefont
  {Santori}, \citenamefont {Solomon}, \citenamefont {Pelton},\ and\
  \citenamefont {Yamamoto}}]{san02}%
  \BibitemOpen
  \bibfield  {author} {\bibinfo {author} {\bibfnamefont {C.}~\bibnamefont
  {Santori}}, \bibinfo {author} {\bibfnamefont {G.~S.}\ \bibnamefont
  {Solomon}}, \bibinfo {author} {\bibfnamefont {M.}~\bibnamefont {Pelton}}, \
  and\ \bibinfo {author} {\bibfnamefont {Y.}~\bibnamefont {Yamamoto}},\ }\Doi
  {10.1103/PhysRevB.65.073310} {\bibfield  {journal} {\bibinfo  {journal}
  {Phys. Rev. B},\ }\textbf {\bibinfo {volume} {65}},\ \bibinfo {pages}
  {073310} (\bibinfo {year} {2002})}\BibitemShut {NoStop}%
\end{thebibliography}

%

\end{document}